\documentclass[10pt,twocolumn,twoside,aps,prl,superscriptaddress,a4paper,nopacs,nokeys,nofootinbib]{revtex4-1}
\usepackage{graphicx}
\usepackage{amsmath,amssymb}
\usepackage{color}
\usepackage{bbm}

\usepackage[colorlinks=true,urlcolor=blue,citecolor=blue,hyperindex,breaklinks=true]{hyperref}

\usepackage{theorem}
\usepackage{latexsym}

\newlength{\blank}
\mathchardef\ordinarycolon\mathcode`\:
\mathcode`\:=\string"8000
\def\vcentcolon{\mathrel{\mathop\ordinarycolon}}
\begingroup \catcode`\:=\active
  \lowercase{\endgroup
  \let :\vcentcolon
  }

\newcommand{\nc}{\newcommand}
\nc{\rnc}{\renewcommand}
\nc{\beq}{\begin{equation}}
\nc{\eeq}{{\end{equation}}}
\nc{\beqa}{\begin{eqnarray}}
\nc{\eeqa}{\end{eqnarray}}
\nc{\lbar}[1]{\overline{#1}}
\nc{\ketbra}[2]{|#1\rangle\!\langle#2|}
\nc{\proj}[1]{| #1\rangle\!\langle #1 |}
\nc{\avg}[1]{\langle#1\rangle}
\nc{\Rank}{\operatorname{Rank}}
\nc{\smfrac}[2]{\mbox{$\frac{#1}{#2}$}}
\nc{\tr}{\operatorname{Tr}}
\nc{\ox}{\otimes}
\nc{\dg}{\dagger}
\nc{\dn}{\downarrow}
\nc{\cA}{\mathcal{A}}
\nc{\cB}{\mathcal{B}}
\nc{\cC}{\mathcal{C}}
\nc{\cD}{\mathcal{D}}
\nc{\cE}{\mathcal{E}}
\nc{\cF}{\mathcal{F}}
\nc{\cG}{\mathcal{G}}
\nc{\cH}{\mathcal{H}}
\nc{\cI}{\mathcal{I}}
\nc{\cJ}{\mathcal{J}}
\nc{\cK}{\mathcal{K}}
\nc{\cL}{\mathcal{L}}
\nc{\cM}{\mathcal{M}}
\nc{\cN}{\mathcal{N}}
\nc{\cO}{\mathcal{O}}
\nc{\cP}{\mathcal{P}}
\nc{\cR}{\mathcal{R}}
\nc{\cS}{\mathcal{S}}
\nc{\cT}{\mathcal{T}}
\nc{\cX}{\mathcal{X}}
\nc{\cZ}{\mathcal{Z}}
\nc{\csupp}{{\operatorname{csupp}}}
\nc{\qsupp}{{\operatorname{qsupp}}}
\nc{\var}{\operatorname{var}}
\nc{\rar}{\rightarrow}
\nc{\lrar}{\longrightarrow}
\nc{\polylog}{\operatorname{polylog}}
\nc{\1}{{\openone}}
\nc{\id}{{\operatorname{id}}}

\nc{\RR}{{{\mathbb R}}}
\nc{\CC}{{{\mathbb C}}}
\nc{\FF}{{{\mathbb F}}}
\nc{\NN}{{{\mathbb N}}}
\nc{\ZZ}{{{\mathbb Z}}}
\nc{\PP}{{{\mathbb P}}}
\nc{\QQ}{{{\mathbb Q}}}
\nc{\UU}{{{\mathbb U}}}
\nc{\EE}{{{\mathbb E}}}

\nc{\qed}{{$\hfill\Box$}}

\begin{document}

\title{Remarks on ``Further comments on ``Rebuttal of ``Refutation of ``Comment on\protect\\ 
       ``Reply to ``Comments on ``A genuinely natural information measure''\,''\,''\,''\,''\,''\,''}
 
\author{Z. Sommer}
\affiliation{Department Mathematik/Informatik--Abteilung Informatik (CS), 
Universit\"at zu K\"oln, Albertus-Magnus-Platz, D-50923 K\"oln, Germany}

\author{A. Winter}
\email{andreas.winter@uni-koeln.de}
\affiliation{Department Mathematik/Informatik--Abteilung Informatik (CS), 
Universit\"at zu K\"oln, Albertus-Magnus-Platz, D-50923 K\"oln, Germany}
\affiliation{ICREA \&{} F\'{\i}sica Te\`{o}rica: Informaci\'{o} i Fenomens Qu\`{a}ntics, %
Universitat Aut\`{o}noma de Barcelona, ES-08193 Bellaterra (Barcelona), Spain}
\affiliation{QUIRCK--Quantum Information Independent Research Centre Kessenich, Gerhard-Samuel-Stra{\ss}e 14, 53129 Bonn, Germany}

\date{1 April 2026}

\begin{abstract}
It's a bit tedious, but as John Doe and Jean Roe have insisted on offering 
further comments on our comprehensive refutation of the former's already 
tiringly obstinate advances, we feel compelled to review their not even wrong 
opinions once again, hoping to put some sense back into the discourse.
\end{abstract}



\maketitle


\noindent
\textbf{1. Reconstruction of the case.}
In the concluding remarks of \cite{Winter:gnats} by the present second author,
the fundamental question ``What?'' had been raised, and the following answer 
was suggested: informally, 
``[t]he pellet with the poison's in the vessel with the pestle.''
(We don't repeat the formal mathematical statement here and rather refer to \cite{Winter:gnats}.)

Subsequently, J. Doe in his comment \cite{Doe:gnats-comment} traced this 
insight back to G. R. Selda and G. S. Wold \cite[pp.~11-12]{SeldaWold:plazzle}, 
citing additionally their corollary ``The chalice from the palace has 
the brew that is true.'' 

This obviously asked for a reply, published in the same issue of 
\emph{Entropy Increase} \cite{Winter:Doe-reply}, in which the older 
reference was gratefully acknowledged, but at the same time pointed out 
that \cite{SeldaWold:plazzle} not only had the confusing error 
``The palace from the chalice...,'' but that their reasoning, if 
not their entire motivation, were complementary to \cite{Winter:gnats} 
(see also \cite{Sommer:personal}). 

J. Doe inexplicably didn't want to let this otherwise crystal-clear matter 
rest, and returned with a long elaboration of his original comment and, 
quite frankly, minor faults he pretended to have found in our exposition 
and reply \cite{Doe:gnats-comment-2}. His article, embarrassingly published 
in an otherwise respectable journal, is a convolution of distortions and 
(we can only assume: deliberate) misunderstandings, which were laid to 
rest by our robust refutation \cite{SommerWinter:Doe-refutation}. 
Let us ignore the wilful attribution to the present second author of the statement 
``[T]he chalice from the palace has the true that is brew;''\footnote{Indeed, ``...brew 
that is true,'' obviously an innocent typo.} 
similarly ``[T]he pellet with the poisle's in the vessel with the plazzle...''\footnote{Eh, the plazzle with the vlessle. Eh, the the bless...}${}^{,}$\footnote{The vessel with the plozle is the plazzle with the...}${}^{,}$\footnote{Clearly, more typos owed to the pressure of having to respond quickly. Only deliberate malice would misconstrue ulterior motivations behind a trivial mishap like that.}

The most egregious manoeuvre, however, was
the following attempt to shift the narrative to new ground: 

\begin{quote}
 The pellet with the poison's in the flagon with the dragon, 
 the vessel with the pestle has the brew that is true.
 \cite[p.~21]{Doe:gnats-comment-2}
\end{quote}
While not entirely unmotivated by the real-world application of the theory,
and in fact somewhat alluded to in \cite[Ch.~2]{SeldaWold:plazzle}, 
this does not aid the previous debate in any way. 
Be that as it may, this and other attempts in Doe's second comment to muddy the 
waters have been summarily dealt with in \cite{SommerWinter:Doe-refutation}, 
the main point of which was to point out that our prior work had 
already anticipated the above in paraphrased form: 

\begin{quote}
 The pellet with the poison's in the flagon with the dragon, 
 the pestle with the pizzle... the pizzle with the f-- 
 the, the, the viss...\footnote{No, no, the vessel with the pestle has...${}^1$}%
              ${}^{,}$\footnote{The poison's in the dragon with the pestle. (Another evident typo, no need to dwell on it as Doe does with pedantic obstinacy, calling it ``yet another howler from the Winters [sic] paper factory.'' \cite[p,~33]{Doe:gnats-comment-2}.)}
 \cite[p.~3]{Winter:Doe-reply}
\end{quote}
In the interest of brevity we refer the reader to \cite{SommerWinter:Doe-refutation}.

Despite our best efforts, J. Doe felt the need to publish yet another rebuttal 
of our clarifications \cite{Doe:gnats-comment-3}. Since its publication, 
J. Roe has waded into the dispute with even further comments \cite{Roe:further-comments}. 
As both share the same tendency, namely to obliquely deny our paper \cite{Winter:gnats}
and our subsequent explanations any original value, we have decided to 
deal with them in one go. The following excerpts should be sufficient to 
illustrate their technique: 

\begin{quote}
 ``The pellet with the chasley, eh... 
 the pellet with the poison is in the pasley with the chazzle.'' 
 \cite[p.~34]{Doe:gnats-comment-3}
 

 ``The pellet with the poisle is in the flaggle with the chalice.'' 
 \cite[p.~35]{Doe:gnats-comment-3}

 ``...the chazzle is in the poisley with the plellice with the plan-- eh, plaglice.'' 
 \cite[p.~11]{Roe:further-comments}
\end{quote}
We cannot dispute these statements, although we lament the somewhat tendentious 
use of quotation marks, which are bound to mislead an impartial reader.

\bigskip\noindent
\textbf{2. Our issues.} 
Our adversaries continue to ignore our bona fide attempts 
to put the original insights of \cite{Winter:gnats} and \cite{SeldaWold:plazzle}
into more commonly accessible terms, or else pretend that our genuine 
trivial typos require elaborate corrections. In our view, we have merely 
clarified, paraphrased and put into historical context our contributions 
from \cite{Winter:gnats} and \cite{Sommer:personal}. The other side, despite 
the best evidence, disingenuously keeps insinuating plagiarism \cite{Kreisler:Oper}, 
which is both ridiculous \cite{Kaye:Tschaikovski,Kaye:Stanislavsky} 
and unoriginal \cite{Lehrer:Lobachevsky}.

We submit to the scientific readership to adjudicate in this case. 
It may be helpful, though, to recall that on the one hand, 

\begin{quote}
The pellet with the dragon's in the pestle with the poi--;

The dragon with the poisle's in the pestle...
\end{quote}
On the other hand, 

\begin{quote}
The poisley with the plazzle is the plazzle with the ploizle;

[T]he floizle with the flagon is the chalice with the poison.
\end{quote}
Note that all of these are verbatim quotes from \cite{Winter:Doe-reply}, 
repeated in \cite{SommerWinter:Doe-refutation}. They speak for themselves, 
and any further argument (short of taking up arms, in a manner of speaking) 
surely is superfluous.

\bigskip\noindent
\textbf{3. Conclusion.}
It is probably fair to say that we are not quite sure anymore what exactly 
we intended to claim in the original article \cite{Winter:gnats}, nor whether we 
minded quite so much John Doe's dissent (nor indeed whether his opinion 
was in fact that far from ours to start with). Be that as it may, we are 
pretty certain that we were right about something, and that he and the others 
trying to get their foot in were wrong about something, though about what 
exactly becomes harder to say each time we engage. Nevertheless, we feel strongly 
about our disagreement, and we are glad to see that John Doe remains equally 
committed. With his and Jean Roe's next round of misguided complaints essentially 
guaranteed, we can assure them and our rapidly dwindling readership that we shall 
be ready.

\bigskip
\noindent
\textbf{Acknowledgments.}
It gives us immense pleasure to implicate a large number of serious 
people and institutions in this work. In fact, the present contribution 
would not have been possible without the support by the European Commission
QuantERA project ExTRaQT (Spanish MICIN grant no.~PCI2022-132965); by the 
Spanish MICIN (project PID2022-141283NB-I00) with the support of FEDER funds; 
by the Spanish MICIN with funding from European Union NextGenerationEU 
(PRTR-C17.I1) and the Generalitat de Catalunya; by the Spanish MTDFP 
through the QUANTUM ENIA project: Quantum Spain, funded by the European 
Union NextGenerationEU within the framework of the ``Digital Spain 
2026 Agenda''; by the Institute for Advanced Study of the Technical University Munich;
by the Alexander von Humboldt Foundation; 
and by the Swedish Research Council under grant no.~2021-06594 while the second 
author was in residence at Institut Mittag-Leffler in Djursholm, Sweden during 
the spring of 2026.


\end{document}